\documentclass[fleqn,usenatbib]{new_mnras}

\usepackage{newtxtext,newtxmath}

\usepackage[T1]{fontenc}

\DeclareRobustCommand{\VAN}[3]{#2}
\let\VANthebibliography\thebibliography
\def\thebibliography{\DeclareRobustCommand{\VAN}[3]{##3}\VANthebibliography}

\usepackage{graphicx}

\newcommand{\jj}{J0946+1006}

\newcommand{\oiii}{[O\,{\sc iii}]}
\newcommand{\ciii}{C\,{\sc iii}]}

\usepackage{color}

\defcitealias{2014MNRAS.443..969C}{CA14}
\newcommand{\ca}{\citetalias{2014MNRAS.443..969C}}
\defcitealias{2020MNRAS.497.1654C}{CS20}
\newcommand{\cs}{\citetalias{2020MNRAS.497.1654C}}

\title[The Jackpot completed]{
A fully-spectroscopic triple-source-plane lens: the Jackpot completed}

\author[R.J. Smith \& T.E. Collett]{
Russell J. Smith$^1$\thanks{russell.smith@durham.ac.uk} and
Thomas E. Collett$^2$\thanks{E-mail:thomas.collett@port.ac.uk}
\\
$^{1}$Centre for Extragalactic Astronomy, University of Durham, Durham DH1 3LE, United Kingdom\\
$^{2}$Institute of Cosmology and Gravitation, University of Portsmouth, Burnaby Rd, Portsmouth, PO1 3FX, United Kingdom\\
}

\pubyear{2021}

\begin{document}

\label{firstpage}
\pagerange{\pageref{firstpage}--\pageref{lastpage}}
\maketitle

\begin{abstract}
We present a spectroscopic determination of the redshift of the second source in the Jackpot gravitational lens system \jj, for which
only a photometric estimate of $z_{\rm phot}$\,=\,2.41$^{+0.04}_{-0.21}$ has previously been available.  
By visually inspecting an archival VLT X-Shooter observation, we located a single emission line from the source in the H-band. 
Among the possible options we find that 
this line is most likely to be \oiii\,5007\,\AA\ at $z_{\rm spec} $\,=\,2.035. 
Guided by this proposal, we were able to detect the faint \ciii\,1907,1909\,\AA\ emission doublet in a deep VLT MUSE datacube. The \ciii\ emission is spatially coincident with the brightest parts of the second Einstein ring, and 
strongly supports the redshift identification.
The spectroscopic redshift is only marginally consistent with the photometric estimate.  Re-examining the cosmological constraints from \jj, the revised measurement favours less negative values of the dark energy equation-of-state parameter $w$;
when combined with a cosmic microwave background prior, we infer $w$\,=\,--1.04$\pm$0.20.
The revised redshift does not significantly help to reconcile the small discrepancy in the image positions for the even more distant third source in \jj.
\end{abstract}


\section{Introduction}

In double-source-plane gravitational lenses, a foreground galaxy forms multiple images of two background objects, at different redshifts.  
Such systems can be used to determine ratios of angular diameter distances, which in turn can be used to constrain cosmological parameters, such as the dark-energy equation-of-state parameter, $w$ \citep[e.g.][]{2012MNRAS.424.2864C}.

The Jackpot lens \jj\ is the best studied example of a galaxy-scale multiple-source-plane system to date. The lens was initially identified through the presence of emission lines at  $z_{\rm s1}$\,=\,0.609 in the Sloan Digital Sky Survey  
spectrum of a $z_{\rm l}$\,=\,0.222 early-type galaxy target \citep{2008ApJ...682..964B}. 
Follow-up imaging with {\it Hubble Space Telescope} (HST) confirmed that this source, hereafter s1, was lensed, forming two bright arcs. The HST data also revealed a second set of arcs at larger radius and hence arising from a more distant source, s2 \citep{2008ApJ...677.1046G}.
Most recently in \citet[][hereafter \cs]{2020MNRAS.497.1654C}, we reported a third source, s3, a doubly-imaged 
Lyman-$\alpha$ emitter at $z_{\rm s3}$\,=\,5.975. 
This source was discovered in a deep integral-field data cube acquired
with the Multi Unit Spectroscopic Explorer 
\citep[MUSE;][]{2010SPIE.7735E..08B} 
on the 
ESO Very Large Telescope (VLT).

While s1 and s3 were identified through their spectroscopic signatures, s2 was detected in
continuum imaging,  
and its redshift was not known from discovery.
\cite{2012ApJ...752..163S}
acquired a spectrum with the Keck telescope
centred on the brightest of the s2 arcs, and covering the wavelength range 3500--8600\,\AA, 
but they were not
able to identify any spectral features from this source.  Analysing HST data in five filters from U to H, they reported a photometric redshift of $z_{\rm phot,s2}$\,=\,$2.41^{+0.04}_{-0.21}$.  

The photometric redshift estimate has been used to derive constraints in the ($\Omega_{\rm m}$, $w$) 
cosmological parameter plane by
\citet[][hereafter \ca]{2014MNRAS.443..969C},
and has been used to test the mass density profile of the main lens by \cite{2012ApJ...752..163S}.
Since these analyses rely critically on the accuracy of the photometric redshift it is of interest to corroborate the value spectroscopically.  (A precise spectroscopic redshift will also improve the precision of these analyses, although the photometric redshift uncertainty is a sub-dominant component of the total error budget for both applications.) With this goal, in \cs, we 
searched the deep optical 
datacube for spectral features from s2,
but found
no convincing lines or continuum breaks.
As noted there, if the published photometric redshift is broadly correct, the MUSE data 
cover rest wavelengths 1400--2800\,\AA, where no strong features are expected. Weaker lines, such as 
[C\,{\sc iii}] at 1907\,\AA\ and C\,{\sc iii}] at 1909\,\AA\ 
(collectively \ciii\ hereafter) might be present, but  could not be identified when working ``blind'', i.e. without advance clues as to the precise redshift.

In this paper,
we establish a secure spectroscopic redshift determination for s2
using archival data from X-Shooter \citep{2011A&A...536A.105V} at the VLT, in combination with the 
\cs\ MUSE observation.
The treatment of the X-shooter data, and identification of 
a probable \oiii\ emission line is described in 
Section~\ref{sec:xshoo}.  Using this line as a guide, Section~\ref{sec:muse} shows the MUSE detection of the faint \ciii\ lines at the same redshift, and co-located with the s2 arc system. 
Section~\ref{sec:obsrev} 
reviews the observational results in the context of prior studies, 
while the implications for previous lensing analyses of \jj\ 
are addressed in Section~\ref{sec:implic}. 
Our conclusions are summarized in Section~\ref{sec:concs}.

\section{A  tentative redshift  from X-SHOOTER}\label{sec:xshoo}

\begin{figure*}
\includegraphics[width=180mm]{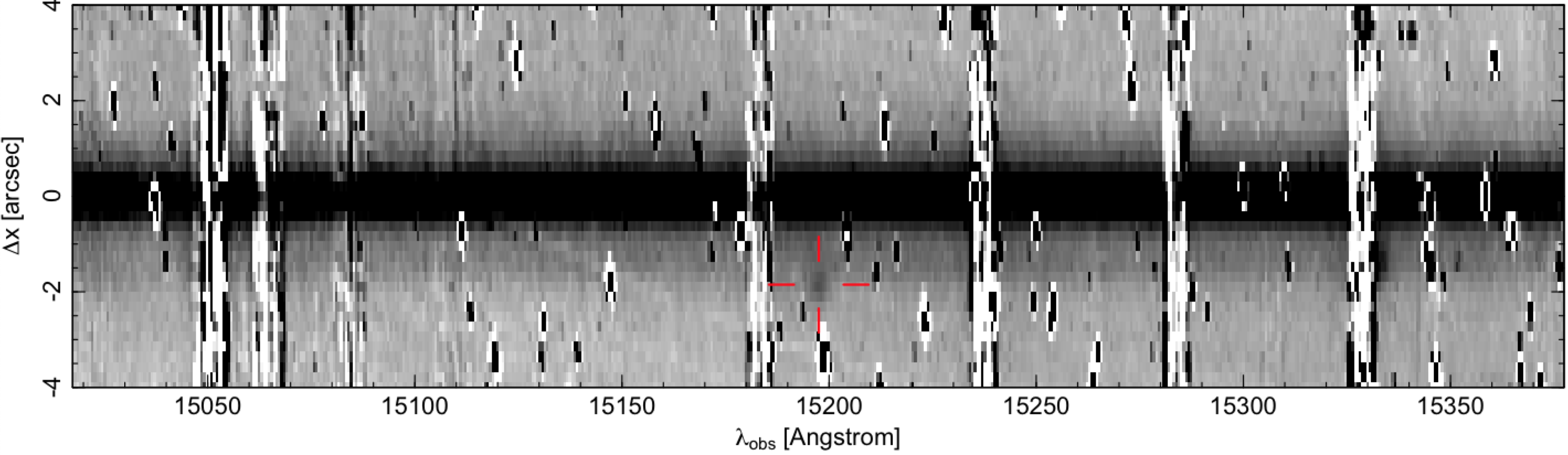}
\vskip -2mm
\caption{Extract from the X-Shooter NIR spectrum, showing the only emission line confidently detected from the s2 arc system (red crosshairs). The strong trace at $\Delta$$x$\,=\,0 is from the lens galaxy, and the faint extension to  $\Delta$$x$\,$\approx$\,--1.5\,arcsec is continuum from the bright s1 arc. As well as the expected residuals from subtracting the bright  H-band sky lines, the spectrum is contaminated by numerous bad pixels and artifacts because the observation was acquired without any dither between exposures. The emission feature at (15335\,\AA, --1.4\,arcsec), partially obscured by sky residuals, is the [S\,{\sc iii}]\,9531\,\AA\ line from s1 at $z$\,=\,0.609.}
\label{fig:xshooline}
\end{figure*}

\begin{figure*}
\includegraphics[width=180mm]{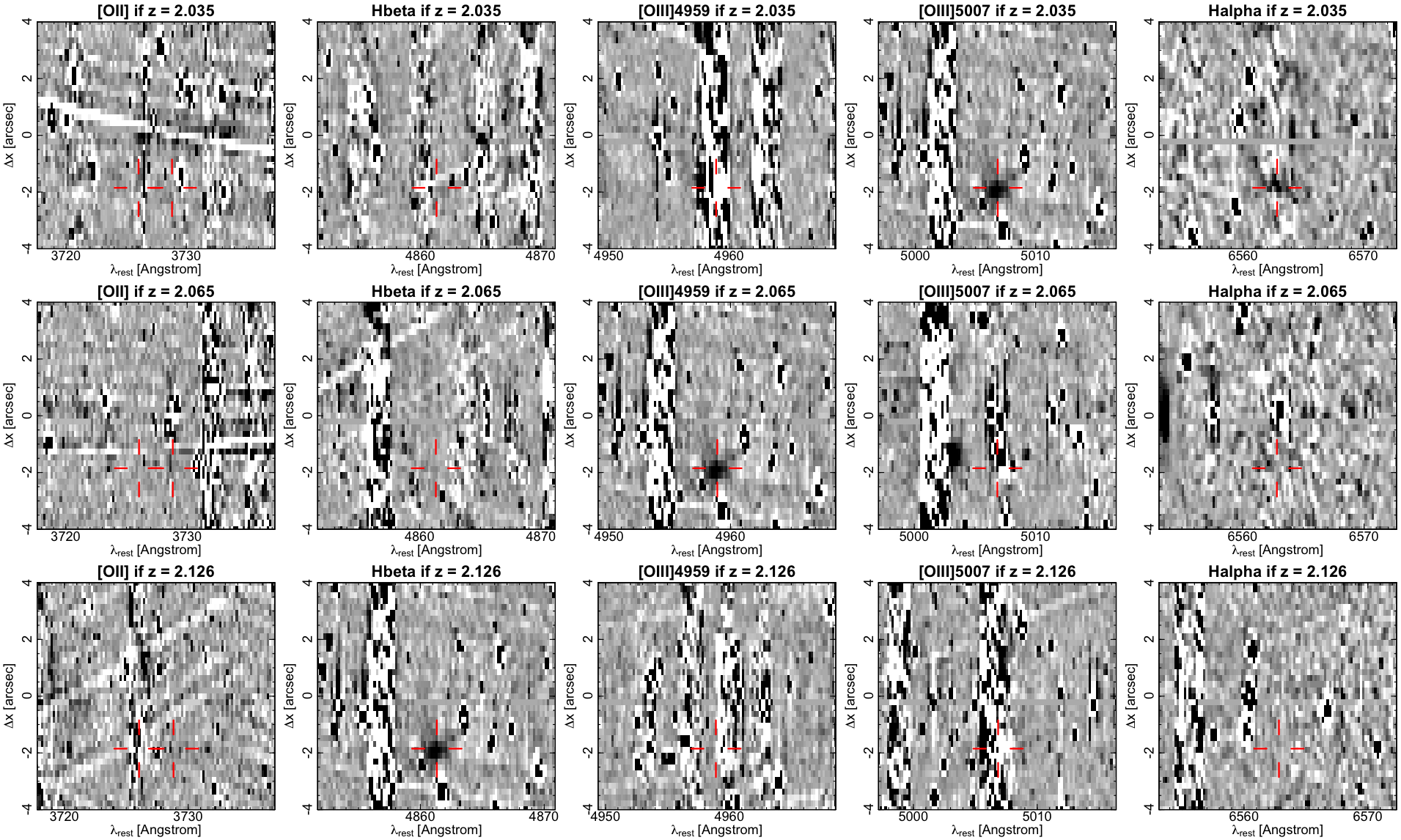}
\vskip -2mm
\caption{A matrix of tests for possible corroborating lines in the X-Shooter spectrum, given different possible identifications for the line detected at 15197.5\,\AA. The top row shows extracts assuming the detected line is [O\,{\sc iii}]\,5007\,\AA,
the row second is the equivalent for 
[O\,{\sc iii}]\,4959\,\AA, and the third is for H$\beta$. The average continuum profile of the lens galaxy has been subtracted to aid visibility. An approximate telluric absorption correction was imposed derived from the lens continuum.
Red markers show the expected locations of emission lines. None of the three possibilities can be confirmed or definitively rejected from these tests.} 
\label{fig:ztestmat}
\end{figure*}

As a first step towards a redshift measurement, we make use of 
archival X-Shooter near-infrared (NIR) spectroscopic data
acquired in 2013 by \cite{2015MNRAS.452.2434S}. 
These observations were taken with a 11$\times$1.2\,arcsec$^2$ slit\footnote{\cite{2015MNRAS.452.2434S} report that a 1.5\,arcsec slit was used, but this option was removed from the X-Shooter NIR arm in 2011.}
at a position angle of 10.3\,degrees, which intersects the s2 Einstein ring close to the faint southern arc (see Discussion below).

The observation comprises 15 exposures totalling 3675\,s. 
All of the frames were taken at an identical sky position,  without any dithering or nodding along the slit, which precludes the automated extraction of a clean 1D spectrum, given the numerous cosmetic defects and artifacts in the X-Shooter NIR detector. Instead, we reduced the data using the standard ESO pipeline, and then 
visually searched the co-added 2D spectrum for features
which could be plausibly attributed to s2. 
In particular, we take into account the known 
separation from the lens centre, and the expected
spatial and spectral line profile. 

Through this method we identified a single emission line in the H-band, at 15197.5\,\AA, offset south along the slit by $\sim$2\,arcsec  (see Figure~\ref{fig:xshooline}). This part of the spectrum is clear of sky residuals and not significantly contaminated by detector artifacts.
The line has spectral FWHM $\sim$\,3.8\,\AA\ (cf. nominal instrumental resolution 3.5\,\AA) and spatial FWHM 0.66\,arcsec, consistent with the nominal image quality. The integrated signal-to-noise ratio is $\sim$20.

Among the plausible identifications for the 15197.5\,\AA\ line, 
[O\,{\sc iii}]\,5007\,\AA\ at $z$\,=\,2.035 
[O\,{\sc iii}]\,4959\,\AA\ at $z$\,=\,2.065, or
H$\beta$ at $z$\,=\,2.126, 
would all be marginally consistent with the \cite{2012ApJ...752..163S} photometric estimate. We examine each of these possibilities in detail in Figure~\ref{fig:ztestmat}.
In the [O\,{\sc iii}]\,5007\,\AA\ at $z$\,=\,2.035 solution, the companion 4959\,\AA\ line is unfortunately coincident with a strong sky subtraction
residual, and cannot provide confirmation. No convincing emission is seen at the corresponding [O\,{\sc ii}], H$\alpha$ or H$\beta$ wavelengths either, but these lines are not necessarily tightly linked to [O\,{\sc iii}], so the solution cannot be excluded.
The [O\,{\sc iii}]\,4959\,\AA\ at $z$\,=\,2.065 solution is disfavoured 
because the (threefold brighter) 5007\,\AA\ companion line is not detected. (There is some contamination from bad pixels in this part of the spectrum, but such a strong line would likely be discernible.) 
The H$\beta$ at $z$\,=\,2.126 solution is also disfavoured, in this case by the absence of a corresponding (brighter) H$\alpha$ line.

Other redshift solutions would be inconsistent with the photometric estimate, and are again disfavoured by the absence of neighbouring lines. For example, 
[O\,{\sc ii}] 3727\,\AA\ at $z$\,=\,3.077 can be excluded since the  doublet profile would be well resolved by X-Shooter.
If the line is H$\alpha$ at $z$\,=\,1.316 (or [N\,{\sc ii}] 6584\,\AA\ at $z$\,=\,1.308), we would expect to have detected [O\,{\sc ii}] in the MUSE spectrum. 

We conclude that emission line shown in 
Figure~\ref{fig:xshooline} is most likely to be 
[O\,{\sc iii}]\,5007\,\AA, though this cannot be confirmed from the X-Shooter data alone.

\section{Confirmation via the C\,{\sc iii}] doublet in MUSE}\label{sec:muse}

The tentative X-Shooter line identification leads to a candidate redshift of $z$\,=\,2.035. 
Guided by this information, we reanalysed the
\cs\ MUSE observation (5.2\,h exposure, 0.5\,arcsec image quality),
searching for the weak 
\ciii\ lines, which are often the only detectable features in the MUSE spectral range for galaxies at $z$\,=\,1.5--3.0 \citep{2017A&A...608A...4M}. 
The conditions for \ciii\ emission may be expected to trace those required for strong
\oiii, and for a high \oiii/[O\,{\sc ii}] ratio \citep{2016ApJ...833..136J}, as apparently present in s2.

\begin{figure}
\includegraphics[width=83mm]{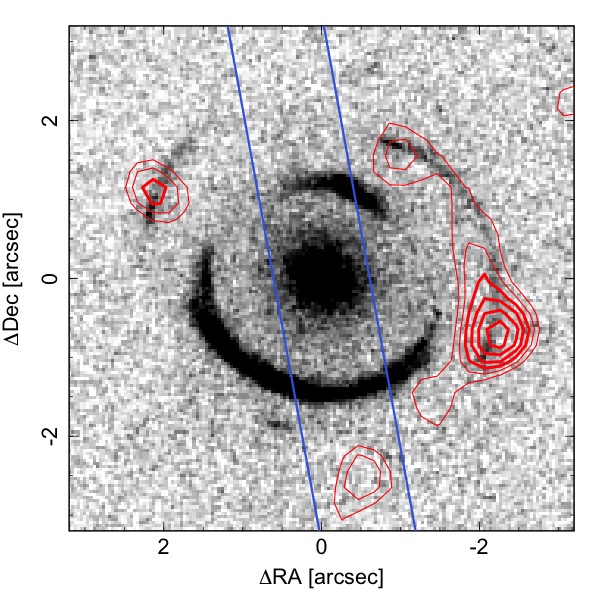}
\includegraphics[width=83mm]{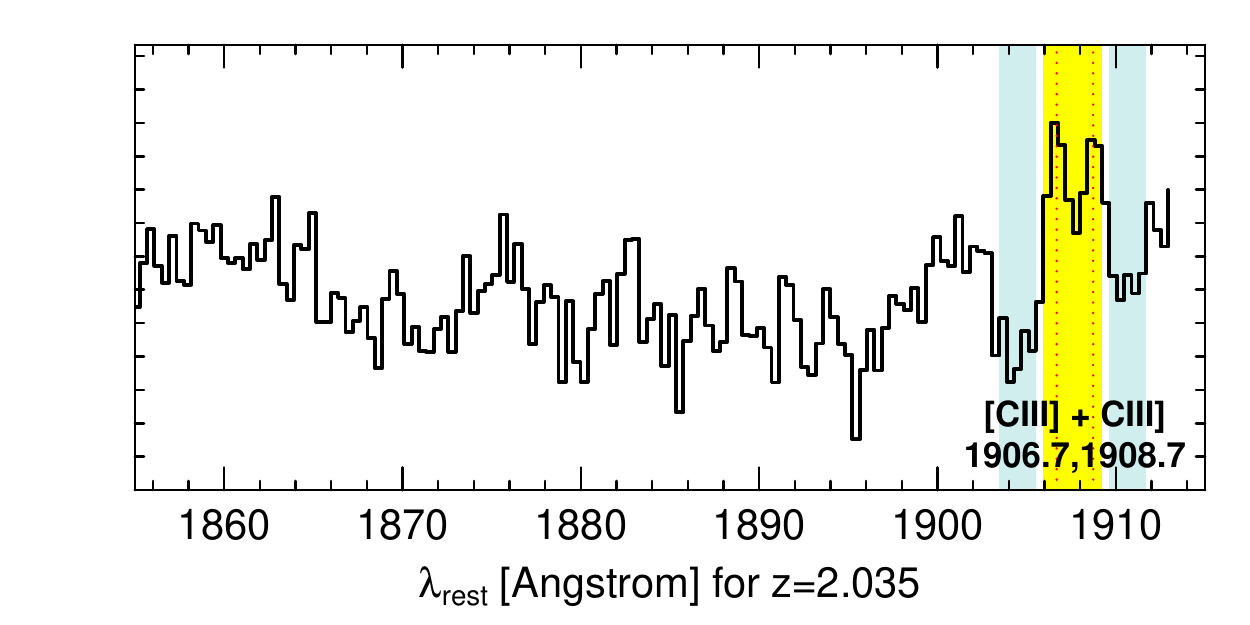}
\caption{The weak C\,{\sc iii}] doublet at $z$\,=\,2.035 recovered in the MUSE data, guided by the X-Shooter evidence. The upper panel shows the HST F438W image in grey scale, with contours from the net narrow-band C\,{\sc iii}] image from MUSE overlaid in red.
Contours are spaced by 0.5$\sigma$ in the excess flux, with the first bold contour drawn at 
2.5$\sigma$, after smoothing with a gaussian of 0.6\,arcsec FWHM.
The blue lines indicate the 1.2\,\AA\ slit used
for the X-Shooter observation.
The lower panel shows the spectrum extracted 
from the pixels inside the bold contour,
with the doublet lines indicated. Yellow and blue regions show the on- and off-band wavelength ranges used for the narrow-band image.
Pixels at $\lambda_{\rm rest}$\,$>$\,1913\,\AA\ are missing due to the MUSE Na\,D rejection filter.}
\label{fig:hstciii}
\end{figure}

Constructing net narrow-band images centred on the expected C\,{\sc iii}] wavelength
indeed reveals a weak feature coincident with the brightest parts of the s2 arcs, 
with a peak of $\sim$4\,$\sigma$ above the noise level. 
The upper panel of Figure~\ref{fig:hstciii} shows the net emission contours overlaid on the HST image, after some tuning of the line and continuum bands. 
The line emission matches closely to the 
s2 arc configuration, with peaks near both ends of the merged-image western arc, where the continuum is also strongest, and at the corresponding point in the north-east arc.
We do not detect \ciii\ emission from the 
fainter southern arc, however.
(No \ciii\ emission is detected here in the X-Shooter spectrum either.) 
Extracting a spectrum from the brightest pixels in the narrow-band image reveals a doublet structure consistent with correct line separation to be C\,{\sc iii}].
The redshift places the lines very close to the region excised by the Na\,D notch filter used in MUSE to reject light from the 
adaptive optics (AO) laser guide sources. For the alternative redshift solutions of $z$\,=\,2.065 and $z$\,=\,2.126, 
C\,{\sc iii}] would fall within the missing  wavelength range, so the equivalent test cannot be performed for these cases.
By extracting similar narrow-band images 
with randomly-selected central wavelengths, we find 
that spurious ``detections'' are fairly frequently observed somewhere in the field, as expected given the low signal level. However, comparably close matches to the arc morphology are not generated by chance.

The C\,{\sc iii}] emission could not have been (and was not) reliably detected from the MUSE datacube alone, without 
the  precise guidance from the X-Shooter line detection (see Section~\ref{sec:obsrev}). 
Taking the 
two datasets in combination, however, the concordant
wavelengths of [O\,{\sc iii}] and C\,{\sc iii}], and the spatial coincidence of C\,{\sc iii}] with the s2 arcs, leave little doubt that the second source in the \jj\ lens system lies at $z_{\rm s2}$\,=\,2.035.

\section{Observational discussion}\label{sec:obsrev}

As shown in Figure~\ref{fig:hstciii}, the 
X-Shooter observation was made with the slit oriented at position angle +10.3\,deg,
such as to not intersect with the s2 arc system
as seen in the HST continuum image. 
Thus it may seem surprising that any line emission was detected from s2 at all.\footnote{Our search for lines in the X-Shooter spectrum was in fact made before correctly establishing the position angle. The image header records the angle with opposite sign to the slit angle on the sky. Not realising this, we believed the slit had been oriented at --10.3\,deg which would have intersected both the southern arc and the north tip of the long western arc.
Another  X-Shooter observation (programme 090.B-0834 with PI Gavazzi) was taken in the IFU mode at position angle +25\,deg, and the 4$\times$1.8\,arcsec$^2$ field of view again does not intersect the s2 arcs at all. 
} 
The most straightforward explanation 
is that the observed \oiii\ flux was carried into the slit by convolution with the $\sim$0.6\,arcsec ground-based seeing. If most of the line emission even from the southern arc was missed by X-Shooter, the full s2 arc system must be very bright in \oiii.
Alternatively, some of the line emission may arise in parts of the source that are not coincident with the UV continuum.
A future AO-assisted near-infrared IFU observation, e.g. with the Enhanced Resolution Imager and Spectrograph (ERIS) at the VLT \citep{2018SPIE10702E..09D}, would be able to map the \oiii\ emission throughout the s2 arcs, providing further lensing information and also independent dynamical constraints on the mass 
in the $z$\,=\,2.035 plane.\footnote{Such an observation would simultaneously map the [S\,{\sc iii}]\,9069,\,9531\,\AA\ lines from s1, which are both clearly visible in the H-band with X-Shooter.}

The apparently clear detection of  \ciii\ in this paper contrasts with the results of \cs, where we were unable to find any emission from s2 in the same MUSE datacube.
Although the coincidence of the emission with the HST arc morphology seems compelling, the contours shown in Figure~\ref{fig:hstciii} do reflect some adjustment of the line and continuum bandpasses and spatial smoothing scale. This treatment is only justified with the prior knowledge of the probable redshift, derived from the X-Shooter line.  An equivalent degree of hand-tuning in a search over all possible redshifts, even if practical, would likely have generated many spurious matches.  Note that the alternative approach, also tried by \cs, of extracting a net spectrum from all pixels which sample the arcs, yields a weaker average signal than that shown in Figure~\ref{fig:hstciii}, because the line emission is concentrated towards one end of the source.

Our spectroscopic measurement of $z_{\rm s2}$ is only marginally consistent ($\sim$2$\sigma$) with the photometric estimate derived by \cite{2012ApJ...752..163S}. 
They summarize the redshift constraint as 
$z_{\rm phot,s2}$\,=\,$2.41^{+0.04}_{-0.21}$, and the full posterior probability distribution (their figure~4) is quite asymmetric, with a low-redshift tail extending to $z$\,=\,1.5
and a small secondary peak at $z$\,=\,1.75.
The photometric redshift was based on images in five filters,
of which the bluest are F336W and F438W.  
The redshift ambiguity may arise partially from the gap in coverage
between these two filters at 3600--4000\,\AA,
corresponding to a Ly$\alpha$ continuum break at $z$\,=\,2.0--2.3.
We note also that the Ly\,$\alpha$ emission line is redshifted to 3688\,\AA\ at $z$\,=\,2.035, which is just within the wavelength range covered by the
3.5-hr integration Keck spectrum of \cite{2012ApJ...752..163S}. No line is visible at that wavelength in their published 1D spectrum, nor in a stacked 2D spectrum re-reduced from their raw data. The \cite{2015MNRAS.452.2434S} X-Shooter UVB-arm data also show no detectable Ly\,$\alpha$ emission.

\section{Implications of the revised redshift}\label{sec:implic}

In this section, we consider the implications of the revised redshift for some previous analyses of the \jj\ lens system.

The deflection angle for rays passing the primary lens of \jj\ is a function only of their impact position through the lens plane. However, two rays passing the same position in the lens plane trace to different unlensed angular positions if they originate from different source planes. The reduced deflection angles (the vector on the sky between observed and unlensed positions) scale as $D_{\rm l,s}/D_{\rm s}$, where 
the $D$ are angular diameter distances to the source
from the lens and from the observer, respectively.
It is this scaling effect that results in the concentric Einstein rings seen in \jj: the more distant source produces a larger ring. In the limit of an isothermal lens and a massless first source, the ratio of the Einstein ring radii is given by the cosmological scaling factor $\beta_{\mathrm{CA14}}$ \footnote{The definition of $\beta$ changes as more sources are added. For ease of comparison with \ca\ we have used the two-source definition here, ignoring the third source discovered in \cs.}:
\begin{equation}
\label{eq:beta}
\beta_{\mathrm{CA14}} \equiv \frac{D_{\rm l,s1} D_{\rm s2}}{D_{\rm s1} D_{\rm l,s2}}.
\end{equation}
By performing a lens reconstruction of the arcs in \jj\, with an elliptical power-law density profile, plus an external shear, and allowing for mass on the first (s1) source plane, \ca\ measured $\beta_{\mathrm{CA14}}$\,=\,0.712$\pm$0.008. This  reconstruction was performed entirely in angular coordinates: redshifts only enter when mapping the measurement of $\beta_{\mathrm{CA14}}$ into constraints on the cosmological parameters.

\begin{figure}
\includegraphics[width=86mm]{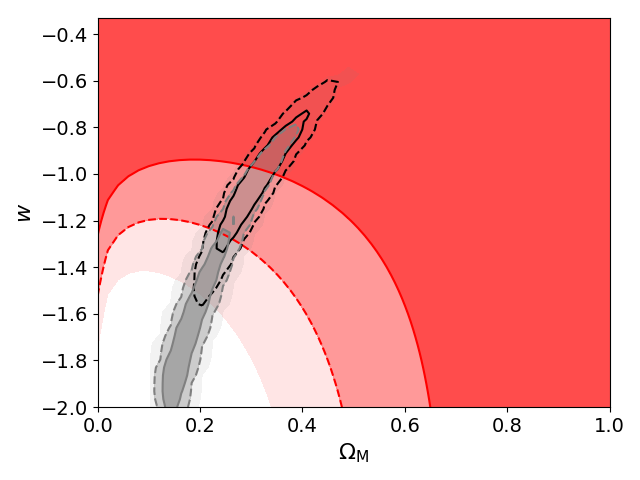} 
\vskip -3mm
\caption{The $w$ and $\Omega_M$ plane. The red contours show the 68, 95 and 99.7 per cent confidence regions derived from the \ca\ measurement of the cosmological scale factor in \jj\ and the updated redshift for s2. 
The {\it Planck} 2018 constraints are shown in grey, while black contours show the combined inference from \jj\ and {\it Planck}, with the solid (dashed) line encircling the combined 68 (95) percent confidence region.}
\label{fig:cosmoplot}
\end{figure}

Assuming a flat $w$CDM cosmology, where the dark energy equation of state is constant but not fixed to the $\Lambda$CDM value of $-1$, \ca\ found $w$\,=\,--1.17$\pm$0.20 when
$\beta_{\mathrm{CA14}}$
is combined with a prior from  {\it Planck} measurements of the Cosmic Microwave Background (CMB) \citep{2014A&A...571A..16P}.
With the updated redshift for s2, the measured value of $\beta_{\rm CA14}$ maps to a less negative value of $w$ at a given $\Omega_{\rm m}$. Combined with the final {\it Planck} constraints \citep{2020A&A...641A...6P},
the result now becomes $w$\,=\,--1.04$\pm$0.20. We show the two-dimensional $(\Omega_{\rm m},w)$ parameter space in Figure \ref{fig:cosmoplot}. As in \ca, cosmologies with very low matter density and very negative equation of state are strongly excluded by the \jj\ result. These solutions are preferred by {\it Planck} alone, but the combination of \jj\ with the CMB data prefers a region of parameter space consistent with flat $\Lambda$CDM.

The newly measured redshift also affects the consistency test for the s3 image positions that we presented in \cs\ (figure 3 therein). We showed that the best fitting lens model from \ca\ maps the two images 
of the $z$\,=\,5.975 galaxy back to positions which differ in the source plane by 0.15\,arcsec. This is larger than the estimated astrometric errors, but can be accounted for by adjusting the most influential model parameters within their 1$\sigma$ errors. In particular,
the s3 image positions can be reconciled either by 
flattening
the primary lens mass density profile, or by increasing the value of $\beta_{\mathrm{CA14}}$. The new redshift of s2 does not change the value of $\beta_{\mathrm{CA14}}$ inferred from the lens model, only its mapping to the cosmological parameters. 
However, for a lower $z_{\rm s2}$, 
a larger mass of s1 is necessary to produce the same lensing effect on s2. This in turn increases its deflection of rays from s3 by five percent. Since lensing by s1 is 
only a perturbation in the \jj\ system, this effect only brings the delensed image positions in the s3 plane 0.02\,arcsec closer together than before. Thus the new redshift for s2  does not 
change the conclusions of \cs: the best fit lens model from \ca\ is a good but not perfect description of the lensing of s3. A full 
four-plane lens model is still required to  exploit fully the lensing information of all three sources
(D. Ballard et al., in preparation).

Finally, we recall that \cite{2012ApJ...752..163S} used the \jj\ system  to constrain the stellar and dark-matter mass distribution in the lens galaxy itself. 
In this work, the cosmology is assumed to be known, so the change in $z_{\rm s2}$ 
enters directly through decreasing $D_{\rm l,s2}/D_{\rm s2}$ 
by 1.7 per cent compared to their adopted $z_{\rm s2}$\,=\,2.4. To balance this, the lensing mass within the second ring is increased by 1.7\,per cent, which will lead to a very slightly flatter inferred total mass profile. This change is only marginally significant for the analysis of \cite{2012ApJ...752..163S} since their uncertain Einstein radius measurement already maps to a 2 per cent uncertainty on the mass within the second ring. 
Calculating the projected masses with the ingredients of their decomposed model, we find such a shift could be generated by reducing the stellar fraction by $\sim$10\,per cent for fixed dark-matter
slope, or by flattening the halo profile from $\gamma$\,=\,1.7 to 1.6, at fixed fractional contributions. Due to the large degeneracies inherent to the dark and stellar mass decomposition, changing the redshift of s2 to 2.035 thus has negligible impact on the 1D marginalised results of \cite{2012ApJ...752..163S}.

\section{Conclusions}\label{sec:concs}

We have used X-Shooter and MUSE observations in combination to establish a secure spectroscopic redshift for the second source in the Jackpot lens \jj. 
The revised redshift
shifts the inferred value of the dark energy equation-of-state parameter to less negative values than derived in previous work. Combined with the {\it Planck} CMB prior, and marginalising over $\Omega_{\rm m}$, 
we find $w$\,=\,--1.04$\pm$0.20. Whilst the combined result is in good agreement with $\Lambda$CDM there is a weak tension between the two constraints, with 
the lens analysis favouring smaller $|w|$ than the CMB.

The Jackpot is now a fully spectroscopic triple-source-plane lens system, making it a high-precision optical bench for cosmology. 
A future comprehensive analysis of \jj\ will incorporate simultaneous modelling of all three sources including the compound lensing effects, as well as complementary dynamical modelling of the spatially-resolved stellar kinematics from MUSE.

\section*{Acknowledgements}
RJS was supported by the Science and Technology Facilities Council through the Durham Astronomy Consolidated Grant 2020--2023 (ST/T000244/1). TEC was supported by the Royal Society through a University Research Fellowship.
This research is based on observations collected at the European Organisation for Astronomical Research in the Southern Hemisphere under ESO programmes 
089.A-0364(A) and 0102.A-0950. This research has made use of the Keck Observatory Archive (KOA),
  which is operated by the W. M. Keck Observatory and the NASA
  Exoplanet Science Institute (NExScI), under contract with the
  National Aeronautics and Space Administration.

\section*{Data availability}
The observational data used in this paper are publicly available in the ESO (\url{archive.eso.org}), HST (\url{hla.stsci.edu}) and Keck (\url{www2.keck.hawaii.edu/koa/public/koa.php})
archives.

\bibliographystyle{mnras}
\bibliography{s2frag}

\bsp	
\label{lastpage}
\end{document}